\documentclass[twocolumn,letterpaper,10pt]{IEEEtran}
\usepackage{cite}
\usepackage{amsmath,amssymb,amsfonts}
\usepackage{algorithmic}
\usepackage{multirow}
\usepackage{array}
\usepackage{tabularx}
\usepackage{booktabs}
\usepackage{graphicx}
\usepackage{textcomp}

\begin{document}
\title{Accurate Airway Tree Segmentation in CT Scans via Anatomy-aware Multi-class Segmentation and Topology-guided Iterative Learning}
\author{Puyang Wang, Dazhou Guo, Dandan Zheng, Minghui Zhang, Haogang Yu, Xin Sun, Jia Ge, Yun Gu, Le Lu~\IEEEmembership{Fellow,~IEEE}, Xianghua Ye and Dakai Jin~\IEEEmembership{Member,~IEEE}
\thanks{
P. Wang, D. Guo, L. Lu, and D. Jin are with Alibaba Group (email: wangpuyang.wpy@alibaba-inc.com). 
D. Zheng, H. Yu, X. Sun, J. Ge and X. Ye are with The First Affiliated Hospital of Zhejiang University.
M. Zhang and Y. Gu are with Shanghai Jiao Tong University.
}
}

\maketitle

\begin{abstract}
Intrathoracic airway segmentation in computed tomography (CT) is a prerequisite for various respiratory disease analyses such as chronic obstructive pulmonary disease (COPD), asthma and lung cancer. Due to the low imaging contrast and noises execrated at peripheral branches, the topological-complexity and the intra-class imbalance of airway tree, it remains challenging for deep learning-based methods to segment the complete airway tree (on extracting deeper branches). Unlike other organs with simpler shapes or topology, the airway’s complex tree structure imposes an unbearable burden to generate the ``ground truth” label (up to 7 or 3 hours of manual or semi-automatic annotation on each case). Most of the existing airway datasets are incompletely labeled/annotated, thus limiting the completeness of computer-segmented airway. In this paper, we propose a new anatomy-aware multi-class airway segmentation method enhanced by topology-guided iterative self-learning. Based on the natural airway anatomy, we formulate a simple yet highly effective anatomy-aware multi-class segmentation task to intuitively handle the severe intra-class imbalance of the airway. To solve the incomplete labeling issue, we propose a tailored self-iterative learning scheme to segment toward the complete airway tree. For generating pseudo-labels to achieve higher sensitivity (while retaining similar specificity), we introduce a novel breakage attention map and design a topology-guided pseudo-label refinement method by iteratively connecting breaking branches commonly existed from initial pseudo-labels. Extensive experiments have been conducted on four datasets including two public challenges. The proposed method ranked 1st in both EXACT’09 challenge using average score and ATM’22 challenge on weighted average score. In a public BAS dataset and a private lung cancer dataset, our method significantly improves previous leading approaches by extracting at least (absolute) 7.5\% more detected tree length and 4.0\% more tree branches, while maintaining similar precision.

\end{abstract}



\section{Introduction}
\label{sec:introduction}
Respiratory diseases impose an immense health burden worldwide as millions of people die each year because of chronic obstructive pulmonary disease (COPD), asthma, lung cancer, etc~\cite{iyer2016quantitative,kesimer2017airway,levine2022global}. As one of the major organs in the respiratory system, the intrathoracic airway has been studied broadly for disease screening, diagnosis, surgical navigation, and treatment effect evaluation~\cite{Kuwano1993,smith2018human}. Computed tomography (CT) based quantitative lung analysis provides very valuable diagnostic information, where airway tree segmentation is often a prerequisite for various downstream tasks. Because of the complexity of tree topology in three dimensional (3D) space and hundreds of slices to be annotated, manual airway segmentation is an extremely tedious process costing up to $7~$hours per patient case~\cite{Tschirren2009}.

\begin{figure*}[!t]
\centering
\includegraphics[width=0.85\textwidth]{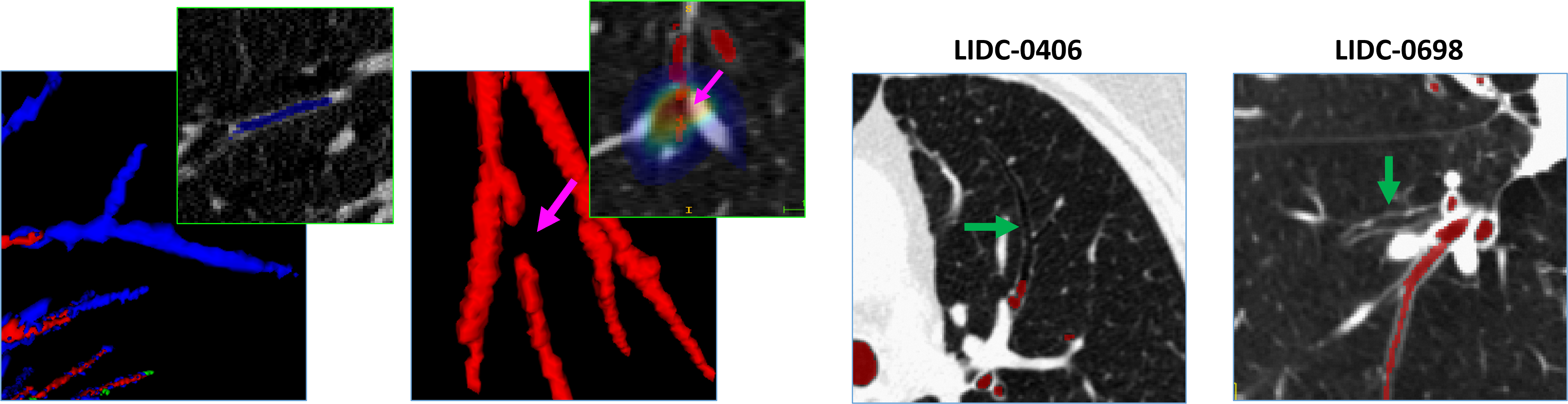}\\
 (a) Missing/Breaking Predicted Branch  \hspace{8em} (b) Incomplete Reference Label 
\caption{Examples of airway tree segmentation with missing and breaking branches (a) produced using the leading method~\cite{zheng2021refined}. Missing annotations are colored in blue and the segmentation breakage is indicated by pink arrow. Note that our proposed breakage attention map can highlight the disconnecting regions (overlaid with CT slice of breaking branch shown in top-right zoomed figure of (a)). (b) shows that the airway tree is often incompletely labeled in the public dataset (e.g., BAS~\cite{qin2019airwaynet}).  } \label{Fig:airwaySeg_demo}
\end{figure*}

Automatic airway segmentation in CT has been extensively explored for decades. Early work involve thresholding, morphology, graph or conventional learning based methods~\cite{Tschirren2005,van2009automatic,Lo2010,pu2012ct,van2013automated,breitenreicher2013hierarchical,bauer2014graph,Xu2015}, where they focus on extracting deeper airway branches while avoiding large segmentation leakages. However, these works often have limited performance with the detected tree lengths, e.g., $<65\%$~\cite{lo2012extraction}. Recent deep learning based approaches achieve significantly improved performance~\cite{Charbonnier2017,jin20173d,yun2019improvement,garcia2019joint,selvan2020graph,qin2019airwaynet,qin2020learning,zheng2021alleviating,zheng2021refined} as reported.  When equipped with 3D deep networks (exploiting the 3D continuity of tree branches), large segmentation leakages can be significantly reduced and efforts has been put to improve the sensitivity and completeness of the segmented airway tree. A 3D fully convolutional network (FCN) is trained followed by a fuzzy connectivity-based segmentation refinement \cite{jin20173d}. Graph neural network is explored to incorporate features of neighborhood for airway segmentation~\cite{selvan2020graph}. Segmentation task is formulated as connectivity prediction~\cite{qin2019airwaynet} and further extends to extract more peripheral branches by feature recalibration and attention distillation~\cite{qin2020learning,qin2021learning}. Another line of research lies in designing new loss functions to enhance airway's tubular connectivity or leverage the inter- and intra-class imbalance problem. For example, a radial distance loss is presented to increase/complete airway's topology ~\cite{wang2019tubular}, and a distance-weighted Tversky loss (named general union loss) is introduced to resolve the intra-class imbalance problem ~\cite{zheng2021alleviating}. Alleviating the airway gradient erosion and dilation problem based on the quantification of local class imbalance in the loss function is reported in ~\cite{zheng2021refined}.  

Despite the boosted performance by deep learning based approaches, there are challenges leading to unsatisfactory segmentation results: (1) By nature, airway possesses a tree topology with locally elongated branches, which have large scale and context variations at different branch levels. For example, diameters of trachea and small bronchus can be differed by 10 to 20 times where locations and surrounding tissues are also different (mediastinum vs. lung parenchyma). Substantial intra-class imbalance and context differences increase the learning difficulty especially for those thin branches at peripheral locations. Although various strategies are developed~\cite{wang2019tubular,zheng2021alleviating,zheng2021refined,qin2021learning}, false negative branches (missing branches) can still be commonly observed in the leading methods (Fig.~\ref{Fig:airwaySeg_demo}(a)). (2) Due to high uncertainty in distal bronchi that yield poor contrast between airway lumen and wall, the presence of disconnected branches, i.e., segmentation breakage, is almost unavoidable (Fig.~\ref{Fig:airwaySeg_demo}(a)). Connecting the breakage would benefit the sensitivity and completeness of the extracted airway tree. Recent work~\cite{shit2021cldice} has investigated new topology-preserving loss (clDice \cite{shit2021cldice}) to enforce the segmentation connectivity, but it requires a significantly longer training time and has a markedly inferior performance as compared to other state-of-the-art airway segmentation methods~\cite{qin2020learning,zheng2021refined,zheng2021alleviating}. (3) Unlike other organs with simpler shape or topology, airway's complex tree structure imposes an unbearable burden to generate the ``ground truth" label, as there are often tens of various airway branches appearing a 2D slice with hundreds of slices to be annotated. Manual or semi-automatic annotation can lead to 7 or 3 hours to fully label one case~\cite{Tschirren2009}.  Therefore, most airway datasets are incompletely labeled (Fig.~\ref{Fig:airwaySeg_demo}(b)). For instance, EXACT'09 challenge simply generated the reference label of the testing set by ensembling true airway branches from the predictions of participant teams~\cite{lo2012extraction}. Training on these incomplete airway reference labels would limit the resulted segmentation performance.

In this work, we propose a new anatomy-aware multi-class segmentation method enhanced by topology-guided iterative self learning to tackle the airway segmentation challenges. Motivated by its tree structure, a simple yet highly effective anatomy-aware multi-class segmentation (AMC) is formulated to intuitively handle severe intra-class imbalance and large context variation of branches at different tree branch levels. Specifically, we decompose the entire airway tree based on the anatomic levels~\cite{tschirren2005matching} into 3 classes and directly learn the class-specific features, i.e., trachea and left/right main bronchi (large branches mostly at mediastinum), bronchi up to the segmental level, and the rest of bronchi visible in CT (small branches deep into lung parenchyma). As shown in the experiments, this effective strategy significantly boosts the performance with 2.9\% $\sim$ 5.1\% more detected tree length. Additionally we incorporate a general union loss (GUL)~\cite{zheng2021alleviating} to further alleviate the severe intra-class imbalance problem in airway segmentation.

\begin{figure*}[!t]
\centering
\includegraphics[width=0.88\textwidth]{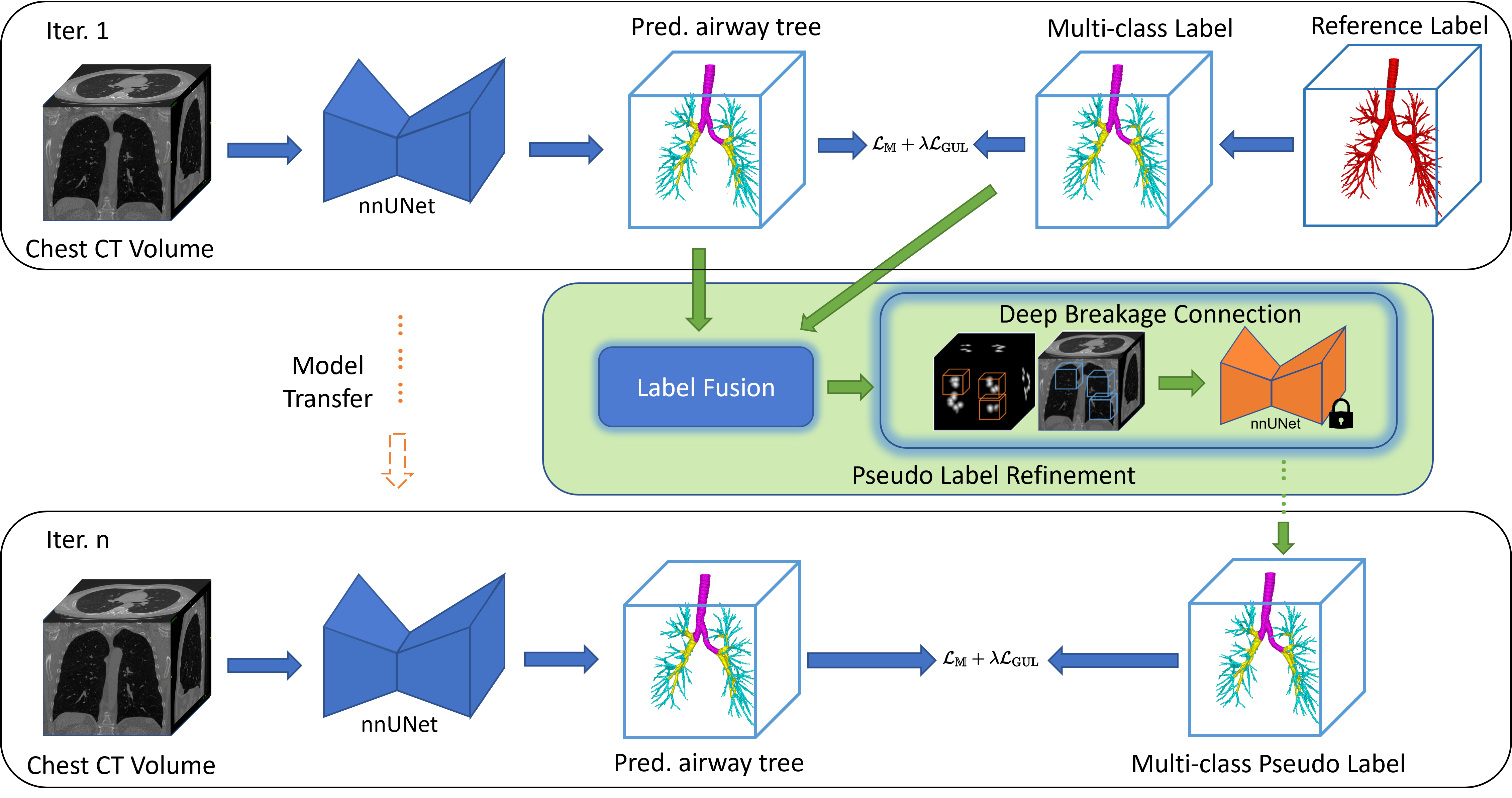}
\caption{Overall workflow of our proposed breakage reducing airway segmentation framework named DeepTree. Note that for the inference and training of Deep Breakage Connection module, only patches around the breakage attention are used as inputs to the network.}\label{Fig:airwaySeg_overall_pipeline} \vspace{-2mm}
\end{figure*}

It has been observed that deep networks properly trained on incomplete reference labels can extract more or deeper ``true" airway branches (not included in the training reference label)~\cite{jin20173d}. Motivated by this, we propose a tailored self-iterative learning scheme with topology-guided pseudo-label refinement to segment towards a complete airway tree. The key factor for effective iterative learning lies in the pseudo-label quality used in each learning step. For airway, it implies that training labels for the next learning step $n$ should contain more/deeper ``true" airway branches that were previously missed by training labels used in the learning step $n-1$. A straightforward way to generate a refined pseudo-label in step $n$ is to keep the largest component of the predicted airway in step $n-1$, which removes all predictions disconnected from the main airway tree. This operation is highly effective to reduce the false positive segmentation, which otherwise impairs the training in the next iteration. However, it also removes potential ``true" branches that are disconnected from the main airway tree (Fig.~\ref{Fig:airwaySeg_demo}(a)) limiting the sensitivity of pseudo-labels. To solve this, we first introduce a breakage attention map (highlighting the breaking area in disconnected branches), which is used to develop a breakage-connection deep network that connects these breaking branches in the predicted airway label. After connecting disconnected ``true" branches, the largest connected component can be calculated to serve as refined pseudo-labels for the next iteration training. This guarantee pseudo-labels of higher sensitivity (while retaining the similar specificity) to be used in the next iterative learning step.

We extensively evaluate the proposed method on four airway segmentation datasets of various lung diseases, disease severities, and imaging protocols (including two airway segmentation challenges: EXACT’09 \cite{lo2012extraction} and ATM’22 \cite{zhang2023multi}). The proposed method ranks 1st in EXACT’09 challenge \cite{lo2012extraction} when averaging the reported metrics of tree length detected (TLD), branch detected (BD), and precision. It also ranks 1st in  ATM’22 challenge \cite{zhang2023multi} by using the weighted average score calculated between TLD, BD, precision and Dice score. Tested in Binary Airway Segmentation (BAS) \cite{qin2019airwaynet} and another private lung cancer datasets, our proposed method achieves significant improvements of at least 7.5\% and 4.0\% more TLD and BD, respectively, as compared to previous state-of-the-art approaches such as~\cite{garcia2019joint,qin2020learning,zheng2021refined,zheng2021alleviating,qin2021learning,shit2021cldice}.

\section{Methods}
Fig.~\ref{Fig:airwaySeg_overall_pipeline} depicts an overview of our proposed airway segmentation training framework. During the training process, it iteratively updates the deep segmentation model using original labels and refined pseudo-labels. In each iteration, AMC combined with GUL is adopted to train the deep model. In the pseudo-label refinement module, label fusion is conducted followed by a deep breakage-connection model to generate pseudo-labels of higher sensitivity while retraining the specificity. For inference, the segmentation model with balanced performance identified during the iterative training is used to generate the initial airway segmentation, where the prediction is further processed by the deep breakage-connection network to generate the final segmentation. The following sections provide a detailed description of each major module, including anatomy-aware multi-class segmentation, topology-guided pseudo-label refinement and iterative training. 



\subsection{Anatomy-aware Multi-class (AMC) Segmentation}\label{sec:1st-stage}

\begin{figure}[!t]
\centering
\includegraphics[width=1\columnwidth]{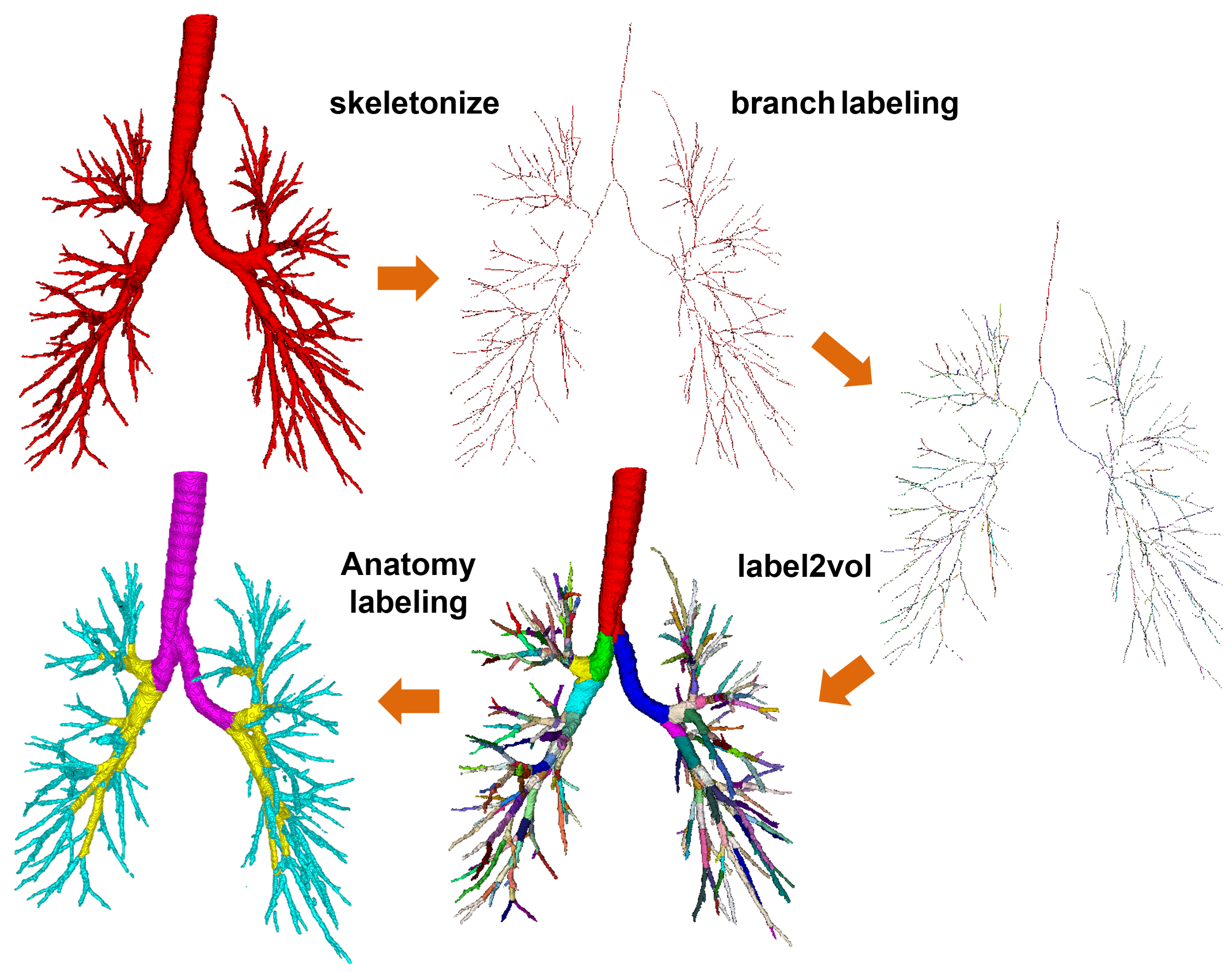}
\caption{Multi-class airway tree label generation using skeletonization and anatomic label matching.}
\label{Fig:airwaySeg_multi_class} \vspace{-3mm}
\end{figure}

As discussed, airway tree structure poses the unique challenge for accurate segmentation and tracing, especially due to the intra-class imbalance between different levels of branches. This could lead to the gradient erosion phenomenon during the network training~\cite{zheng2021alleviating}, making the network ineffective to learn features of small airway branches. To alleviate this issue, we propose a simple and effective strategy by formulating the task to a multi-class segmentation problem, where each class has its distinguished airway branch size range and class-specific features can be more directly learned. This helps to explicitly differentiate the anatomic context of different branches, e.g., trachea in mediastinum and small bronchus in parenchyma. Based on the airway anatomy, we decompose the entire airway tree $Y$ into three classes: (1) trachea $+$ left/right main bronchi $Y^L$, (2) bronchi up to segmental level $Y^M$, and (3) the rest of bronchi $Y^S$. The algorithm flow of generating the multi-class airway labels is illustrated in Fig~\ref{Fig:airwaySeg_multi_class} which consists of the following steps: curve skeletonization~\cite{Jin2016}, curve skeleton labeling~\cite{saha19963d}, skeleton to volume label propagation~\cite{jin2014new}, and multi-atlas based anatomic branch matching\footnote{https://github.com/LiquidFun/Airway}. 



Using the calculated multi-class labels $Y=Y^L\cup Y^M \cup Y^S$, we formulate the anatomy-aware multi-class (AMC) airway segmentation as follows:
\begin{align}
     \mathcal{L}_{\mathbb{M}}(\hat{Y}, Y)=\mathcal{L}_{\mathbb{S}}(\hat{Y}^{L}, Y^{L}) + \mathcal{L}_{\mathbb{S}}(\hat{Y}^{M}, Y^{M}) + \mathcal{L}_{\mathbb{S}}(\hat{Y}^{S}, Y^{S}),
\end{align}
where $\hat{Y}^*$ indicates the predictions and $\mathcal{L}_{\mathbb{S}}$ is the segmentation loss, e.g., Dice loss and/or cross-entropy (CE) loss. Considering that the recent general union loss (GUL)~\cite{zheng2021alleviating} has been shown to be effective on handling intra-class imbalance among sub-categories, we combine it with the proposed AMC loss and apply to the entire airway tree.  
The overall loss function in each iteration is written as follow:
\begin{align}
     \mathcal{L}(\hat{Y}, Y)=\mathcal{L}_{\mathbb{M}}(\hat{Y}, Y) + \lambda \mathcal{L}_{\mathrm{GUL}}(\hat{Y}, Y),
\end{align}
where $\lambda$ is determined empirically. We set $\lambda$ to 0.25 in all our experiments. nnUNet~\cite{isensee2021nnu} is adopted as the backbone to segment the airway tree. Implementation details are described in the Implementation section~\ref{sec:implement}.


\subsection{Topology-guided Pseudo-label Refinement}\label{sec:label-refine}

To generate airway pseudo-labels of higher sensitivity (while retaining similar specificity) used in the next iteration, we design a tailored pseudo-label refinement method by considering the airway tree topology. First, the predicted airway label $\hat{Y}$ in iteration $n$-th is combined with the original reference label $Y$ to get $\hat{Y}'$. This is to add back the missing branches from the original labels (as false negatives) to the predicted $\hat{Y}$. Then, a breakage-connection deep network is utilized to connect breaking branches in $\hat{Y}'$, after which the largest connected-component $\hat{Y}^{cc}$ is extracted to remove the left disconnected predictions from the major tree, which are mainly false positive segmentation. $\hat{Y}^{cc}$ is denoted as the refined pseudo-label to be used in the next iteration of training. Since $\hat{Y}^{cc}$ keeps the same topology as the original airway label, we name this step as topology-guided pseudo-label refinement. The following describes the breakage attention map and how to use it for training the deep breakage-connection network. 

\subsubsection{Breakage Attention Map}\label{sec:attention}
To connect the breaking branches, we first compute a breakage attention map to highlight the breaking area. Given the airway label $\hat{Y}'$, we apply the connected component analysis to label each component $\hat{Y}'_{k}$ leading to a main connected airway tree with the rest disconnected segments. Let $C_k$ denote the number of connected components. Then, for each component $\hat{Y}'_{k}$, a 3D Euclidean distance transformation map $\mathrm{D}_{k}$ is calculated~\cite{maurer2003linear} to provide the shortest distance between a background voxel $j$ and component $\hat{Y}'_k$. Next, let $C_{min}$ denote the component which has the shortest distance to voxel $j$ among all components $\hat{Y}'_{k}, k \in \{1,2,...,C_k\}$.  Then, the raw breakage attention value $H$ at a location $j$ is determined as the second shortest distance among all components:
\begin{align}
H(j) = \underset{k \in \{1,2,...C_k\}, k\neq C_{min} } \min \mathrm{D_{k}}(j)
\end{align}

The raw breakage attention map $H$ can highlight the breaking regions of disconnected branches. To train the breakage-connection deep network, $H$ is further normalized by a Sigmoid function $\bar{H}=\text{Sigmoid}(\gamma -H)$, where $\gamma$ is the parameter for controlling the breakage attention range and we set it to 5 in our experiment. From Fig.~\ref{Fig:airwaySeg_demo}(a), it is observed that $\bar{H}$ at a breakage location normally forms a 3D ball-like intensity distribution.

\subsubsection{Breakage Connection Deep Network}\label{sec:breakage_network}
With the help of breakage attention map, we train a breakage connection deep network to connect these breaking branches. Since breakage happens locally, small local patches are sampled for training the breakage connection network. Specifically, the center of each breakage highlighted in the breakage attention map is first determined. Then, a local CT and breakage attention patch (${X}'$ and $\bar{H}'$) with the size of $64\times64\times64$ is cropped after applying a random 3D spatial shifting. Finally, a 3D breakage connection deep network is trained using the early fusion of $X'$ and $\bar{H}'$ to predict the breakage $\hat{Y}^B$: 
\begin{align}
    \hat{Y}^{B} = \mathcal{F}\left(X', \bar{H}'; \mathbf{W}\right)\\
   \mathcal{L}_{\mathbb{B}} = \mathcal{L}_{\mathbb{S}} \left(\hat{Y}^{B}, Y^{B} \right) ,   \label{eqn:EF}
\end{align}
where $\mathcal{F}(\cdot)$ and $\mathbf{W}$ denote the breakage connection network function and parameters, respectively. $Y^{B}$ and $\hat{Y}^{B}$ indicate the ground truth and predicted breakage, respectively. To generate $\hat{Y}^{cc}$, the outputs $\hat{Y}^{B}$ at all breaking locations are combined with $\hat{Y}'$ and the largest connected component is extracted as the refined pseudo-label for the next iteration training. 

\begin{figure*}[!t]
\centering
\includegraphics[width=0.98\textwidth]{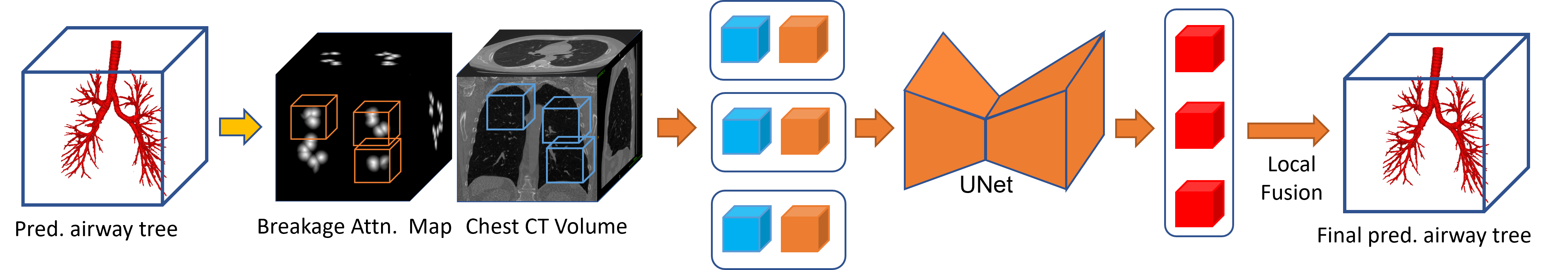}
\caption{The workflow of the proposed breakage connection method, which includes the breakage attention map computation, breakage-connection deep network training, and the local label fusion.}
\label{Fig:airwaySeg_breakage_connection} 
\end{figure*} 

\begin{figure}[!t]
\centering
\includegraphics[width=0.98\columnwidth]{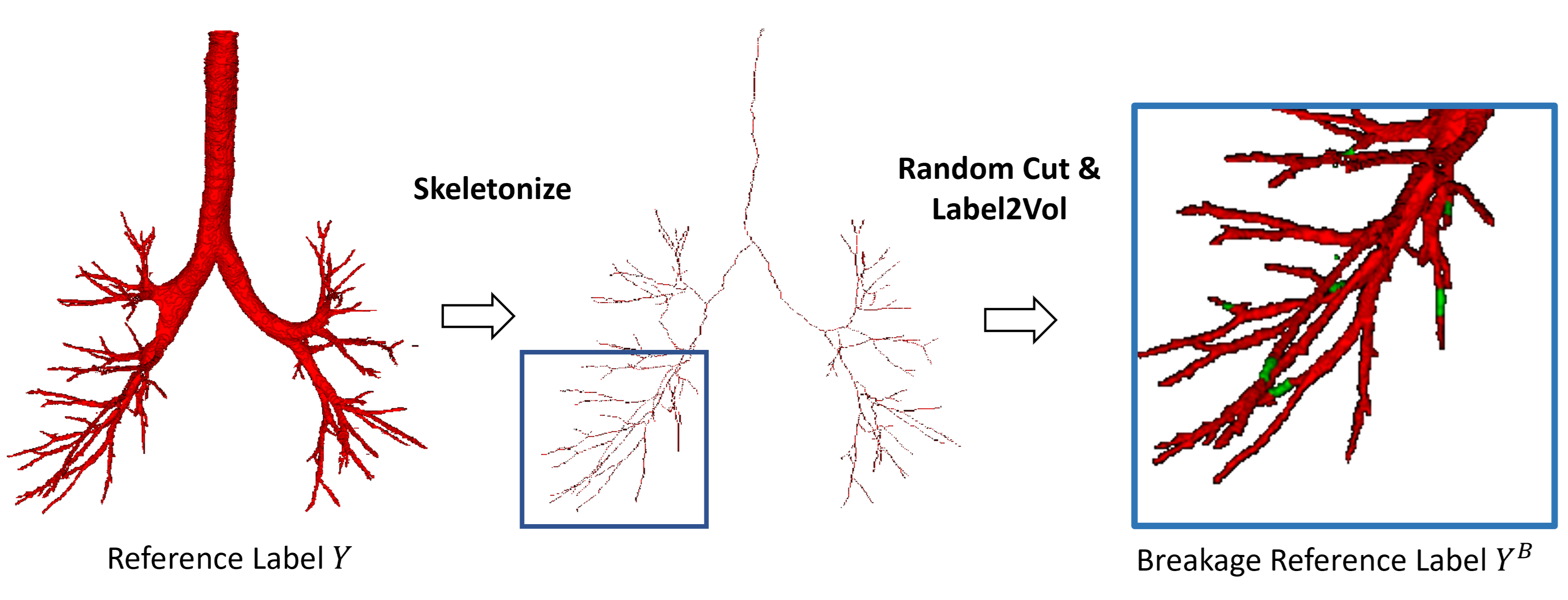}
\caption{An illustration of simulated breakage sample generation. Green color show the simulated breaking branch labels.}
\label{Fig:airwaySeg_breakage_sample_simulate} 
\end{figure}

\subsubsection{Breakage Training Sample Simulation}\label{sec:breakage_sampling}
For training the breakage-connection deep network, we need sufficient training samples that cover various breaking conditions at different airway branch locations. However, there is no breakage in airway reference labels. Therefore, we resort to simulating the branch breaking conditions using reference labels for the breakage-connection training. To achieve this, a domain-specific breakage simulation algorithm is developed as follow: (1) For a reference airway mask in the training set, we first extract its curve skeleton using a robust minimum-cost path approach that avoids generating the false centerlines~\cite{Jin2016}. (2) A random subset $50\%$ of peripheral branches are selected. For a chosen branch, we randomly select a continuous subset of $10\%$-$30\%$ skeleton voxels and set them to the breakage label (different from the airway value). (3) A skeleton-to-volume label propagation algorithm~\cite{Jin2016} is applied to generate the final volumetric breakage ground truth labels ($Y^B$). In this manner, a large number of different breakage conditions can be simulated using the airway reference label. See Fig.~\ref{Fig:airwaySeg_breakage_sample_simulate} for an example. With sufficient training samples, an effective breakage-connection network can be built to connect the breaking branches.



\subsection{Iterative Self-Learning}
As shown in Fig.~\ref{Fig:airwaySeg_overall_pipeline}, initially at the 1st iteration, the multi-class segmentation model is trained using the original airway reference label. Then, the pseudo-label refinement module combines the predicted airway segmentation and original reference label to generate a refined pseudo-label, which is used to update the multi-class segmentation model in the 2nd iteration. CNN weights of the segmentation model in the 2nd iteration are initialized using the pretrained model weights of the 1st iteration. This process is repeated for the maximum of $n$ times with no manual editing involved. Note that for the deep breakage-connection network, it is trained using an initial dataset and subsequently applies to all the experiments without further retraining. Let $X$ and $Y^{n-1}$ denote the CT image and the refined pseudo-label in the iterative learning step $n-1$, respectively. Using $j$ to represent the voxel index, the iterative self learning at step $n$ updates the multi-class airway segmentation model below: 
\begin{align}
\hat{y}^{n}_j&=p_j^{n}\left( y_j^{n-1}=1 | X;  \mathbf{W}^{n}\right) \mathrm{,} \label{eqn:iter-learn} 
\end{align}
where $p_j^{n}(\cdot)$. $\hat{y}^{n}_j$ and $\mathbf{W}^{n}$ denote the CNN functions, output segmentation maps, and corresponding CNN parameters at iteration $n$, respectively, and $y_j^{n-1}$ indicates the refined pseudo-label that is used to update $\mathbf{W}^{n}$.



\section{Datasets and Implementation}

\subsection{Datasets}
Four airway segmentation datasets with a variety of lung diseases, disease severities and imaging protocols are used to comprehensively evaluate the quantitative performance of our proposed airway segmentation method. Specifically, we first train and evaluate our method using Binary Airway Segmentation (BAS) dataset ~\cite{qin2019airwaynet}, which is also utilized to identify the best iteration with balanced performance in self-learning and develop the deep breakage-connection model that both applies to all subsequent datasets for evaluation. BAS dataset also serves in our ablation study to validate the effectiveness of each method component. A private lung cancer dataset is collected as external testing of our model developed using BAS dataset. We participate in two airway segmentation challenges (EXACT'09 and ATM'22) to benchmark and compare our method's performance. Detailed information of each dataset is described below.

\subsubsection{BAS dataset}
We first conduct our experiment on 90 chest CT scans collected in Binary Airway Segmentation (BAS) dataset~\cite{qin2019airwaynet} and follow the same data split in~\cite{zheng2021alleviating} with 50, 20 and 20 cases for training, validation and testing, respectively. BAS dataset contains 70 subjects from the LIDC dataset~\cite{armato2011lung} and 20 cases from the training set of the EXACT’09 dataset~\cite{lo2012extraction}, which are mainly normal subjects or early lung cancer patients.  The in-plane spatial resolution of CT scans ranges from 0.5 to 0.8 mm, and slice thickness varies from 0.5 to 1.0 mm. 
Note that airway reference labels provided in BAS dataset~\cite{qin2019airwaynet} are incomplete, since true airway branches are observed ``being unannotated" as the reference label (see Fig.~\ref{Fig:airwaySeg_demo}(b)). In order to measure the ``real" performance, three board-certified physicians further curate the reference labels in all BAS testing cases by adding the missing true tree branches. We report the segmentation performance using both curated and original reference labels. It worth noting that labels in the training and validation set of BAS are not further curated (i.e., still incomplete in training). We also compare against previous leading methods ~\cite{garcia2019joint,qin2020learning,isensee2021nnu,zheng2021refined,zheng2021alleviating,shit2021cldice} using both curated and original reference labels on this dataset.


\subsubsection{In-house Lung cancer dataset}
We next evaluate the method using a private lung cancer dataset as external testing, consisting of 20 diagnostic CT scans of patients with lung caners. These CT scans of lung cancer patients exhibit different imaging resolutions and various disease severity conditions when compared to BAS dataset. For instance, the in-plane spatial resolution of CT scans ranges from 0.7 to 1.0 mm with slice thickness varied from 1.25 to 2.5 mm. They are patients with lung cancers at more advanced stages in contrast to patients in BAS~\cite{qin2019airwaynet} . Top-performing methods from BAS evaluation are chosen for external comparison, including~\cite{isensee2021nnu,qin2020learning,zheng2021refined,zheng2021alleviating,zhang2022cfda}. The airway labels in this in-house lung cancer dataset are carefully annotated by three physicians, which results in complete airway reference labels. 

\subsubsection{EXACT’09 challenge dataset}
EXACT’09 challenge~\cite{lo2012extraction} provides a training set and a test set each with 20 CT scans with no publicly available training annotation. The in-plane spatial resolution of CT scans ranges from 0.5 mm to 0.8 mm, and slice thickness varied from 0.5 to 1.0 mm. To participate the challenge, we trained our model using the training set with the corresponding annotations from the BAS dataset, and submitted our results on the test set for evaluation. Note that the reference label in EXACT’09 testing set is incomplete, since it is generated by ensembling the true airway branches from predictions of several participant teams~\cite{lo2012extraction,bauer2014graph}.

\subsubsection{ATM’22 challenge dataset}
ATM’22 challenge~\cite{zhang2023multi} contains a large-scale dataset of 500 chest CT scans (300 for training, 50 for validation, and 150 for testing) covering both healthy subjects and patients with severe pulmonary diseases. The in-plane spatial resolution of CT scans ranges from 0.5 to 0.91 mm, and slice thickness varied from 0.5 to 1.0 mm. We trained our model on its training set and submitted our results on the test set for evaluation. According to ATM'22's organizers, the airway reference label in ATM’22 is generated semi-automatically in the following step: each of the 500 CT scan was firstly preprocessed by three segmentation models~\cite{zheng2021alleviating,Cicek2016,yu2022break} developed on the BAS dataset. Then, three predictions were ensembled by majority voting to acquire an initial segmentation result. Theses initial labels were then manually examined and edited by three radiologists using the same annotation principle. Nevertheless we still observe that many reference labels in ATM’22 dataset are incomplete.


\subsection{Implementation Details} \label{sec:implement}
We adopt nnUNet~\cite{isensee2021nnu} as our backbone for both the multi-class segmentation model and the breakage-connection model due to its high accuracy and utility on many reported medical image segmentation problems. We adapt and modify the original nnUNet architecture to better accommodate to the airway task. Specifically, we reduce the downsampling operation to 3 times (originally 5 times) with four levels of spatial resolutions considering the very small branches at peripheral sites. Meanwhile, we enlarge the width of the convolutional layers at the deeper blocks to increase the learning capacity. For the loss function, we implement and use the combination of multi-class Dice + CE loss and GUL for the multi-class segmentation model, while adopting the default Dice + CE loss for the breakage-connection model. The rest training and data augmentation settings are the same as the default nnUNet setup for the initial segmentation model training, e.g., 1000 training epochs with 0.01 learning rate. For the iterative learning step, we update the model with 300 epochs with a 0.001 learning rate, since it is finetuned from the model pretrained in previous steps. For the maximum iteration number, we set it to $5$ in our experiment based on empirical performance observation. To train the breakage connection deep network, local CT and breakage attention patches with the size of 64 × 64 × 64 are cropped around the breakage locations after applying a random 3D spatial shifting. Adaptive nnUNet is adopted and 500 learning epochs is used with the initial learning rate of 0.01.

\subsection{Evaluation Metrics}
For BAS, private lung cancer, and EXACT'09 datasets, we report three most critical and commonly reported airway evaluation metrics~\cite{lo2012extraction,Charbonnier2017,qin2020learning,zheng2021alleviating}: tree length detected (TLD), branch detected (BD) and precision. Note that TLD and BD measure the sensitivity/completeness of the segmented airway tree, while precision indicates inversely the false positive rate of segmentation. For ATM’22 dataset, additional metrics are also used including sensitivity, specificity, Dice score (DSC) and GPU memory consumption~\cite{zhang2023multi}.


\begingroup
\setlength{\tabcolsep}{10pt} 
\renewcommand{\arraystretch}{1.5} 
\begin{table}[!ht]
\caption{Quantitative ablation results for the airway segmentation in BAS dataset. AMC and DBC represent the anatomy-aware multi-class segmentation and the deep breakage connection, respectively. Adapted nnUNet is used as the backbone.}
\label{table_ablation}
\centering
\resizebox{\columnwidth}{!}{%
\begin{tabular}{ccc|ccc}
\hline\hline 
 \multirow{2}{*}{AMC} & \multirow{2}{*}{GUL} &  \multirow{2}{*}{DBC} & \multicolumn{3}{c}{Metrics} \\ \cline{4-6}
          &     &     & BD (\%)$\uparrow$ \quad & TLD (\%) $\uparrow$ \quad         & Precision (\%) $\uparrow$ \\ \hline
 $-$ &     $-$       &     $-$        & $90.3 \pm $5.1    & $78.1 \pm$6.7      &  $\mathbf{99.8 \pm0.2}$ \\
      \checkmark     & $-$ & $-$             & $91.9 \pm $4.7    & $81.0 \pm$5.9      &  $99.4 \pm$0.4 \\
    $-$        & \checkmark    & $-$  & $93.3 \pm $4.4    & $83.1 \pm$5.9      &  $99.0 \pm$0.6 \\ 
  \checkmark &     \checkmark       &  $-$      & $95.1 \pm $3.6    & $88.2 \pm$4.7      &  $98.4 \pm$1.1 \\ \hline
      $-$   & $-$ &        \checkmark     & $91.2 \pm $5.2    & $79.6 \pm$6.5      &  $99.7 \pm$0.3 \\ 
    \checkmark    &        $-$    &  \checkmark & $92.5 \pm $4.5    & $82.3 \pm$5.7      &  $99.4 \pm$0.4 \\ 
   $-$ &     \checkmark       &  \checkmark           & $94.1 \pm $4.0    & $85.9 \pm$5.8      &  $98.9 \pm$0.7 \\
    \checkmark   &    \checkmark    & \checkmark  & $\mathbf{95.9 \pm 3.6}$    & $\mathbf{89.3 \pm4.5}$      &  $98.2 \pm$1.1  \\ \hline \hline
\end{tabular}
}
\end{table}
\endgroup

\begin{figure*}[!t]
\centering
\includegraphics[width=1\textwidth]{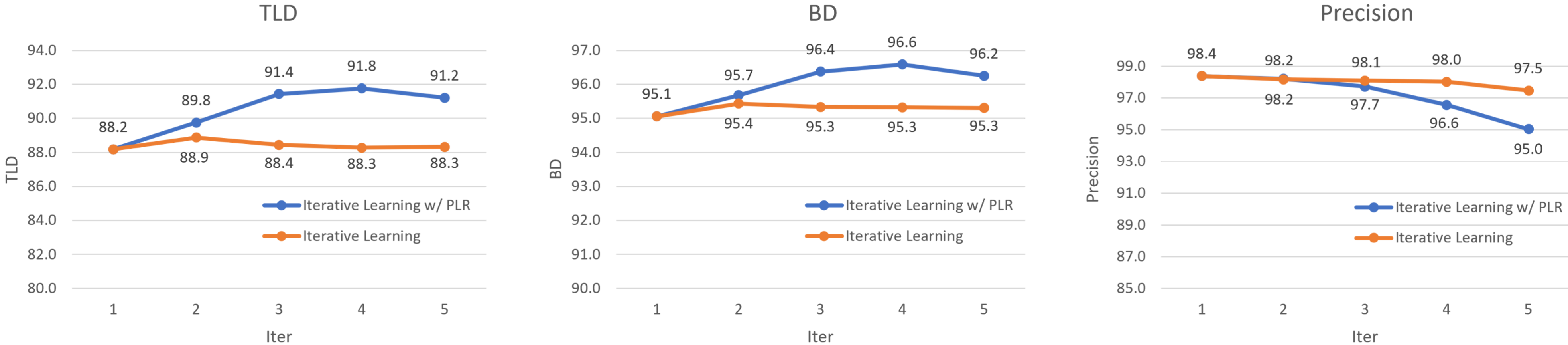}
\caption{Quantitative ablation results for the iterative self-learning of airway segmentation on the BAS dataset. \textit{IL w/ PLR} denotes iterative self-learning with our proposed topology-guided pseudo-label refinement.}
\label{Fig:airwaySeg_iter_ablation}
\end{figure*}

\section{Experimental Results}

\subsection{Evaluation Results on BAS Dataset}

\subsubsection{Comparison with state-of-the-art methods}
The quantitative results (evaluated using both curated and original reference labels in the testing set) and comparisons with state-of-the-art methods are tabulated in Table~\ref{table_bas}. Using the curated (more complete) reference label, it is observed that methods without tailoring to the airway segmentation normally have inferior performance. For example, TLD and BD by UNet++~\cite{zhou2018unet++}, clDice~\cite{shit2021cldice} and nnUNet~\cite{isensee2021nnu} are generally less than those of \cite{zheng2021alleviating,qin2020learning,zheng2021refined} that are specifically designed for airway segmentation. Among the leading airway segmentation methods, two of them produce the previously best performance (80.7\% TLD~\cite{qin2020learning} and 80.1\% TLD~\cite{zheng2021refined}) by using feature recalibration and attention distillation or local-imbalance \& back-propagation based weighting schemes. In comparison, our method (iter=1 and iter=3) significantly improves the performance by {\bf at least 7.5\% and 4.0\% increased TLD and BD}, respectively. It also has the lowest standard derivation. Our precision of 98.4\% at iter=1, 97.7\% at iter=3 is comparable to that of ~\cite{qin2020learning,zheng2021refined} indicating similar false positive results. Two qualitative segmentation examples are shown in Fig.~\ref{Fig:airwaySeg_quality}, illustrating the completeness and robustness of our method.

When evaluated using the reference labels that are originally provided in the testing set, the proposed method achieves $99.1\%$ BD and $98.0\%$ TLD, {\bf at least $1.8\%$ and $4.4\%$ higher} than the second best performed method~\cite{zheng2021refined}. It is observed that the precision calculated using the original reference labels is 86.9\%, however, it drastically increase to at least 97.7\% when examined on the curated reference labels. This shows that the low precision of our method (evaluated using original reference labels) is caused by the incomplete labeling issue instead of the ``real" falsely segmented airway branches. Once the reference label quality is improved, initially considered false positive segmentation is correctly evaluated as true positives. This highlights the importance of reference label curation of the testing set for airway segmentation evaluation. 

\begingroup
\renewcommand{\arraystretch}{1.5} 
\begin{table*}[!t]
\caption{Quantitative comparison with previous state-of-the-art methods in BAS dataset. * denotes results when evaluated using the original reference labels of BAS testing set without further curation.}
\label{table_bas}
\centering
\resizebox{0.77\textwidth}{!}{
\begin{tabular}{l|ccc|ccc}
\hline \hline
\multirow{2}{*}{Methods} & \multicolumn{3}{c|}{Curated Reference Label} & \multicolumn{3}{c}{Original Reference Label} \\
           & BD (\%) $\uparrow$ & TLD (\%) $\uparrow$ & Precision (\%) $\uparrow$  & BD* (\%) $\uparrow$ & TLD* (\%) $\uparrow$ & Precision* (\%) $\uparrow$       \\ \hline
Juarez et al. \cite{garcia2019joint}        & $69.2 \pm$25.4   & $53.5 \pm$20.9      &  $\mathbf{99.9 \pm 0.1}$  & $78.6 \pm$26.4    & $69.1 \pm$25.8      &  $\mathbf{98.1 \pm1.7}$ \\ 
UNet++ \cite{zhou2018unet++}       & $87.2 \pm$10.9    & $74.2 \pm$11.8      &  $99.3 \pm$0.6 & $94.4 \pm$7.6    & $88.7 \pm$10.6      &  $94.3 \pm$2.8 \\ 
clDice \cite{shit2021cldice}  & $88.0 \pm$10.2    & $76.2 \pm$11.1      &  $99.1 \pm$0.7 & $94.5 \pm$7.8    & $90.0 \pm$10.0      &  $93.9 \pm$2.8  \\ 
nnUNet \cite{isensee2021nnu}        & $89.1 \pm$5.5    & $76.1 \pm$6.4      &  $\underline{99.9 \pm0.1}$ & $96.8 \pm$2.7    & $91.7 \pm$5.5      &  $\underline{95.5 \pm 2.5}$   \\
WingsNet \cite{zheng2021alleviating}   & $89.2 \pm$5.8    & $77.1 \pm$5.7      &  $99.0 \pm$0.8   & $95.7 \pm$3.8    & $90.8 \pm$5.6      &  $93.6 \pm$3.1 \\ 
CFDA \cite{zhang2022cfda} & $90.9\pm$6.7    & $78.9\pm$8.1      &  $99.1\pm$0.6 & $97.2\pm$3.3    & $92.7\pm$6.0      &  $92.4\pm$3.3    \\
Qin et al. \cite{qin2020learning}     & $90.9 \pm$8.8    & $80.7 \pm$9.9      &  $98.4 \pm$1.0  & $95.8 \pm$6.1    & $92.4 \pm$8.0      &  $91.5 \pm$3.8 \\
Zheng et al. \cite{zheng2021refined} & $91.1 \pm$5.5    & $80.1 \pm$6.6      &  $98.9 \pm$0.7 & $97.2 \pm$3.0    & $93.4 \pm$4.8      &  $93.2 \pm$3.1    \\ 
\hline
DeepTree (iter=1)\quad & $\underline{95.1 \pm3.6}$    & $\underline{88.2 \pm4.7}$      &  $98.4 \pm $1.1 & $\underline{99.0 \pm1.0}$    & $\underline{97.7 \pm2.3}$      &  $86.9 \pm $4.6  \\
DeepTree (iter=3)\quad & $\mathbf{96.4 \pm3.1}$    & $\mathbf{91.4 \pm3.8}$      &  $97.7 \pm $1.2 & $\mathbf{99.1 \pm1.0}$    & $\mathbf{98.0 \pm2.2}$      &  $84.5 \pm $5.5  \\
\hline \hline
\end{tabular}
}
\end{table*}
\endgroup

\subsubsection{Ablation results}
We conduct the following ablation study using BAS dataset to verify the effectiveness of each module in the proposed method: (1) effectiveness of AMC and GUL in initial airway segmentation; (2) utility of deep breakage-connection in the pseudo-label refinement; (3) performance impact of topology-guided iterative learning. Quantitative ablation study results are illustrated in Table~\ref{table_ablation} and Fig.~\ref{Fig:airwaySeg_iter_ablation}. 

\textbf{Effectiveness of AMC \& GUL:} From the top half of Table~\ref{table_ablation}, it can be observed that the proposed AMC is effective in segmenting the airway tree. When applied to nnUNet  with default Dice + CE loss, AMC improves the performance (row 2 vs. row 1) by obtaining  $\sim$3\% and $\sim$2\% higher TLD and BD, respectively, while approximately maintaining the same high precision. Moreover, when applied to nnUNet with default segmentation loss plus GUL, AMC markedly improves the performance (row 4 vs. row 3) by $\sim5\%$ and $\sim2\%$ increased TLD and BD, respectively. This validates our assumption that decomposing the whole airway tree into 3 classes based on their anatomic levels is of great benefits to handle the intra-class imbalance and large context variations which is intrinsic to airway tree. On the other hand, AMC and GUL are complementary to each other. Although AMC or GUL alone can help better segment the airway, their combination further significantly improves the results (row 4 vs. row 2 \& 3). 

\textbf{Effectiveness of deep breakage-connection:} It is observed from Table~\ref{table_ablation} that our deep breakage-connection model consistently improves all the initial segmentation results (row 5-8 vs. row 1-4). For instance, the detected tree length has been improved by 1.6\% when averaged among four different initial segmentation results with the maximum improvement of 2.8\% (row 7 vs. row 3). Meanwhile, segmentation results refined by the deep breakage-connection model maintain almost the same precision across four settings. These results demonstrate that our breakage-connection method may be universally effective to increase the sensitivity of elongated anatomic structure segmentation, where breaking branches commonly exists~\cite{wu2021scs}.

\textbf{Effectiveness of topology-guided iterative learning:} Fig.~\ref{Fig:airwaySeg_iter_ablation} illustrates the segmentation results of iterative self-learning with five iterations. Beside the proposed topology-guided pseudo-label refinement, we compare to a straightforward iterative learning method, where the pseudo-label is simply refined by keeping the largest connected component in the predicted airway segmentation. With the proposed topology-guided pseudo-label refinement (PLR), it can be seen that self-learning keeps increasing TLD and BD (more previously missed airway branches are extracted) until the forth iteration. At the same time, the precision experiences a slight decrease until the 3rd iteration (from 98.4\% to 97.7\%), after which a markedly drop is observed in iteration 4 (96.6\%) and 5 (95.0\%). Based on these observations, the segmentation model updated after the 3rd iteration provides an overall balanced performance with increased sensitivity and relatively similar precision. Hence, we use the model updated at the third iteration in the later experiments. In comparison, without the topology-guided PLR, self learning is ineffective, as TLD, BD, and precision all stay unchanged across different iterations.


\subsection{Experimental Results on Lung Cancer Dataset}
We use the lung cancer dataset as an external testing set for models developed in BAS dataset. Five top-performing methods in the BAS dataset are compared, and the quantitative results are summarized in Table~\ref{table_lungcancer}.  It is observed that methods generally yield decreased performance when directly tested on the lung cancer dataset, especially in terms of BD and TLD (e.g., $>10\%$ decrease in BD and $>15\%$ decrease in TLD for~\cite{isensee2021nnu,zheng2021refined,qin2020learning,zheng2021alleviating}). Yet, CFDA~\cite{zhang2022cfda} and our method exhibit the least performance drop, i.e., $\sim$9\% decrease in BD. Note that CFDA is a dedicated method aiming to increase the generalization ability in airway segmentation using feature disentanglement and augmentation framework. In contrast, our method does not design complex scheme to improve the generalizability. In terms of precision, methods all show slightly lower values, where precision of our method is marginally higher than that of the second and third best performed methods~\cite{zheng2021refined,zhang2022cfda}. 


\begingroup
\setlength{\tabcolsep}{10pt} 
\renewcommand{\arraystretch}{1.5} 
\begin{table}[!ht]
\caption{Quantitative comparison with previous state-of-the-art methods when externally tested on the lung cancer dataset.}
\label{table_lungcancer}
\centering
\resizebox{\columnwidth}{!}{%
\begin{tabular}{lccc}
\hline Method & BD $(\%) \uparrow$ & TLD $(\%) \uparrow $ & Precision $(\%) \uparrow $ \\
\hline nnUnet \cite{isensee2021nnu}  & $72.6 \pm 8.2$ & $60.2 \pm 7.4$ & $\mathbf{98.8 \pm 1.8}$ \\
WingsNet \cite{zheng2021alleviating}  & $71.7 \pm 9.0$ & $59.8 \pm 7.6$ & $98.4 \pm 1.7$ \\
Qin et al. \cite{qin2020learning}   & $76.4 \pm 10.9$ & $65.4 \pm 9.6$ & $\underline{98.5 \pm 2.6}$ \\
Zheng et al. \cite{zheng2021refined}  & $75.3 \pm 9.6$ & $62.7 \pm 7.1$ & ${97.1 \pm 1.6}$ \\
CFDA \cite{zhang2022cfda} & $\underline{81.9 \pm 10.4}$ & $\underline{68.2 \pm 9.6}$ & $97.4 \pm 3.4$ \\
\hline Proposed & $\mathbf{87.1} \pm \mathbf{7.9}$ & $\mathbf{74.3} \pm \mathbf{8.5}$ & $97.8 \pm 2.0$ \\
\hline
\end{tabular}
}
\end{table}
\endgroup


\begin{figure}[!t]
\centering
\includegraphics[width=1\columnwidth]{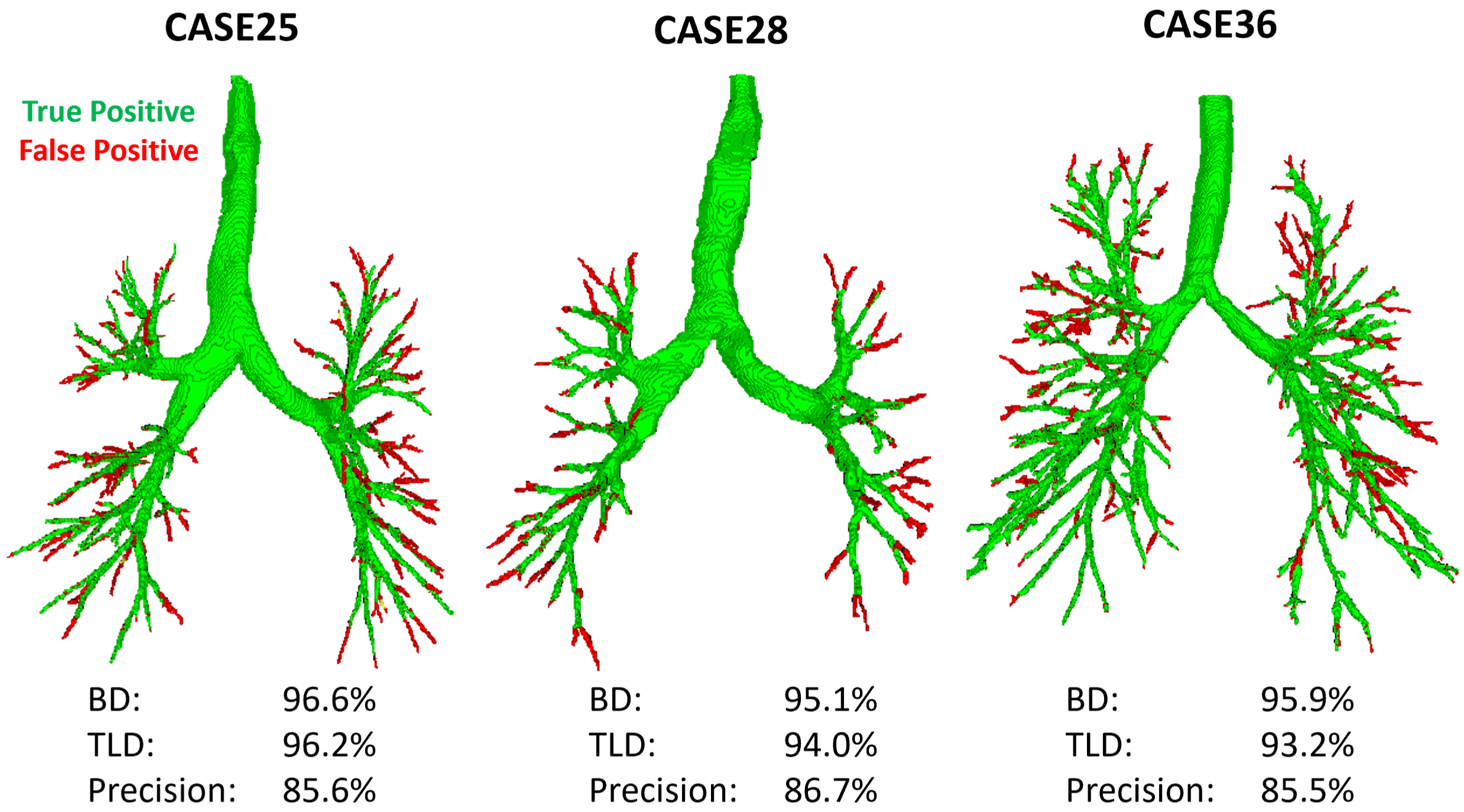}
\caption{Three cases with the lowest precision on the EXACT’09 dataset. Most false positives (shown in red color) belong to unannotated distal branches with no significant leakage observed. 
}
\label{Fig:exact09_quality} \vspace{-3mm}
\end{figure}

\subsection{Comparative Evaluation on EXACT’09 Dataset}
Table~\ref{table_exact09} shows the quantitative results in EXACT’09 dataset. Among the participants of this challenge, the best BD (82.0\%) and TLD (79.4\%) are achieved by Qin et al. \cite{qin2021learning} with a parameter threshold of $0.1$. They adopt a feature recalibration and an attention distillation-based method. Compared with their results, the proposed method achieves superior performance in all three metrics (BD: 86.5\% vs 82.0\%, TLD: 87.1\% vs 79.4\%, precision 91.4\% vs 90.3\%). Note that for the tree completeness measurement of BD and TLD, our method remarkably improves by 4.5\% and 7.7\%, respectively. In comparison, BD and TLD differences among other top three performed methods~\cite{inoue2013robust,zheng2021refined,qin2021learning} are only 2.4\% and 0.9\%, respectively. These quantitative results indicate that our method can extract a significantly more complete airway tree by a large increased margin against previous leading methods. The ``relatively lower" precision of our method is mainly caused by the incomplete reference label in the test set of EXACT'09, previously observed by several work, as Fig. 9 in~\cite{bauer2014graph} and Fig. 5 in~\cite{zheng2021alleviating}. Fig.~\ref{Fig:exact09_quality} shows three testing examples where most false positives (red color) are observed as natural extensions of true airway branches (i.e., unannotated in the original reference label).

\begingroup
\setlength{\tabcolsep}{10pt} 
\renewcommand{\arraystretch}{1.5} 
\begin{table}[!ht]
\caption{Evaluation results on the Exact'09 Challenge test set.}
\label{table_exact09}
\centering
\resizebox{\columnwidth}{!}{
\begin{tabular}{lccc}
\hline Method & BD $(\%)$ $\uparrow$ & TLD $(\%)$ $\uparrow$ & Precision $(\%)$ $\uparrow$ \\
\hline 
Murphy et al. \cite{nardelli2015optimizing}  & $41.6 \pm 9.0$ & $36.5 \pm 7.6$ & $\mathbf{99.3} \pm 1.7$ \\
MISLAB  & $42.9 \pm 9.6$ & $37.5 \pm 7.1$ & $\underline{99.1 \pm 1.6}$ \\
Xu et al. \cite{Xu2015} & $51.1 \pm 10.9$ & $43.9 \pm 9.6$ & $93.2 \pm 26.6$ \\
Bauer et al. \cite{bauer2014graph}& $63.0 \pm 10.4$ & $58.4 \pm 13.2$& $98.6 \pm 2.1$ \\
Feuerstein et al. \cite{feuerstein2009adaptive} & $76.5 \pm 13.3$& $73.3 \pm 13.4$ & $84.4 \pm 9.5$\\
Inoue et al. \cite{inoue2013robust} & $79.6 \pm 13.5$ & $\underline{79.9 \pm 12.1}$ & $88.1 \pm 13.2$ \\
Zheng et al. \cite{zheng2021alleviating}  & ${80.5 \pm 12.5}$ & $79.0 \pm 11.1$ & $94.2 \pm 4.3$ \\
Qin et al. \cite{qin2021learning} $(th = 0.5)$ & $76.7 \pm 11.5$ & $72.7 \pm 11.6$ & $96.3 \pm 2.9$ \\
Qin et al. \cite{qin2021learning} $(th = 0.1)$ & $\underline{82.0 \pm 9.9}$ & $79.4 \pm 10.0$ & $90.3 \pm 5.6$ \\
\hline Proposed & $\mathbf{86.5} \pm \mathbf{9.3}$ & $\mathbf{87.1} \pm \mathbf{6.7}$ & $91.4 \pm 4.5$ \\
\hline
\end{tabular}
}
\end{table} \vspace{-3mm}
\endgroup

\begingroup
\setlength{\tabcolsep}{10pt} 
\renewcommand{\arraystretch}{1.5} 
\begin{table*}[!t]
\caption{Evaluation results on the full hidden test set of ATM’22 challenge.}
\label{table_atm}
\centering
\resizebox{\textwidth}{!}{
\begin{tabular}{clcccccccc}
\hline \hline
Rank & Team / Method & BD $(\%)$ $\uparrow$ & TLD $(\%)$ $\uparrow$ & DSC $(\%)$ $\uparrow$&  Precision $(\%)$ $\uparrow$ & Sen. $(\%)$ $\uparrow$& Spe. $(\%)$ $\uparrow$& GPU memory $\downarrow$ & Weighted Score $\uparrow$\\
1 & \textbf{DeepTree (Ours)} & $\mathbf{97.9 \pm 2.3}$ & $\mathbf{97.1 \pm 3.4}$ & $92.8 \pm 2.2$ & $87.9 \pm 4.2$ & $\mathbf{98.5 \pm 1.4}$ & $99.96 \pm 0.018$ & \textbf{3.52 GB} &  $\mathbf{95.36}$ \\
2 & Timi & $\underline{95.9 \pm 5.2}$ & $\underline{94.7 \pm 6.4}$ & $93.9 \pm 3.7$ & $93.5 \pm 3.4$ & $94.5 \pm 5.2$ & $99.98 \pm 0.012$ & 3.81 GB & \underline{94.85} \\
3 &  YangLab & $94.5 \pm 8.6$ & $91.9 \pm 9.4$ & $\underline{94.8 \pm 7.9}$ & $94.7 \pm 8.3$ & $95.0 \pm 8.2$ & $99.98 \pm 0.010$ &6.24 GB& 93.68\\
4 & neu204 & $90.9 \pm 10.4$ & $86.7 \pm 13.1$ & $94.1 \pm 8.0$ & $93.0 \pm 8.4$ & $\underline{95.3 \pm 8.6}$ & $99.98 \pm 0.013$ &9.00 GB& 90.24\\
5 &  Sanmed AI & $88.8 \pm 7.3$ & $83.4 \pm 10.9$ & $\mathbf{95.0 \pm 1.8}$ & $95.1 \pm 3.2$ & $95.0 \pm 3.3$ & $99.98 \pm 0.011$ &\underline{3.74 GB} &  89.14 \\
6 & dolphins & $90.1 \pm 6.5$ & $84.1 \pm 11.2$ & $92.7 \pm 2.1$ & $94.7 \pm 3.4$ & $91.1 \pm 4.3$ & $99.98 \pm 0.012$ &10.43 GB & 89.13\\
7 & suqi & $89.2 \pm 7.3$ & $82.2 \pm 12.3$ & $93.6 \pm 2.1$ & $95.8 \pm 3.3$ & $91.8 \pm 4.3$ & $99.99 \pm 0.012$ &11.47 GB&  88.39 \\
8 & notbestme & $87.5 \pm 9.0$ & $81.3 \pm 13.6$ & $94.5 \pm 2.3$ & $\mathbf{96.6 \pm 2.7}$ & $92.7 \pm 4.3$ & $\mathbf{99.99 \pm 0.009}$ & 10.58 GB & 87.77 \\
9 & dnai & $86.7 \pm 5.4$ & $77.9 \pm 8.7$ & $90.9 \pm 1.7$ & $91.7 \pm 2.8$ & $90.9 \pm 1.7$ & $99.97 \pm 0.011$ & 11.59 GB & 85.00 \\
10 & lya & $85.2 \pm 9.1$ & $75.7 \pm 14.9$ & $93.8 \pm 2.2$ & $\underline{96.5 \pm 2.9}$ & $91.4 \pm 4.8$ & $\underline{99.99 \pm 0.010}$ &9.58 GB &  84.86 \\
\hline \hline
\end{tabular}
}
\end{table*} \vspace{-3mm}
\endgroup

\begin{figure}[!t]
\centering
\includegraphics[width=1\columnwidth]{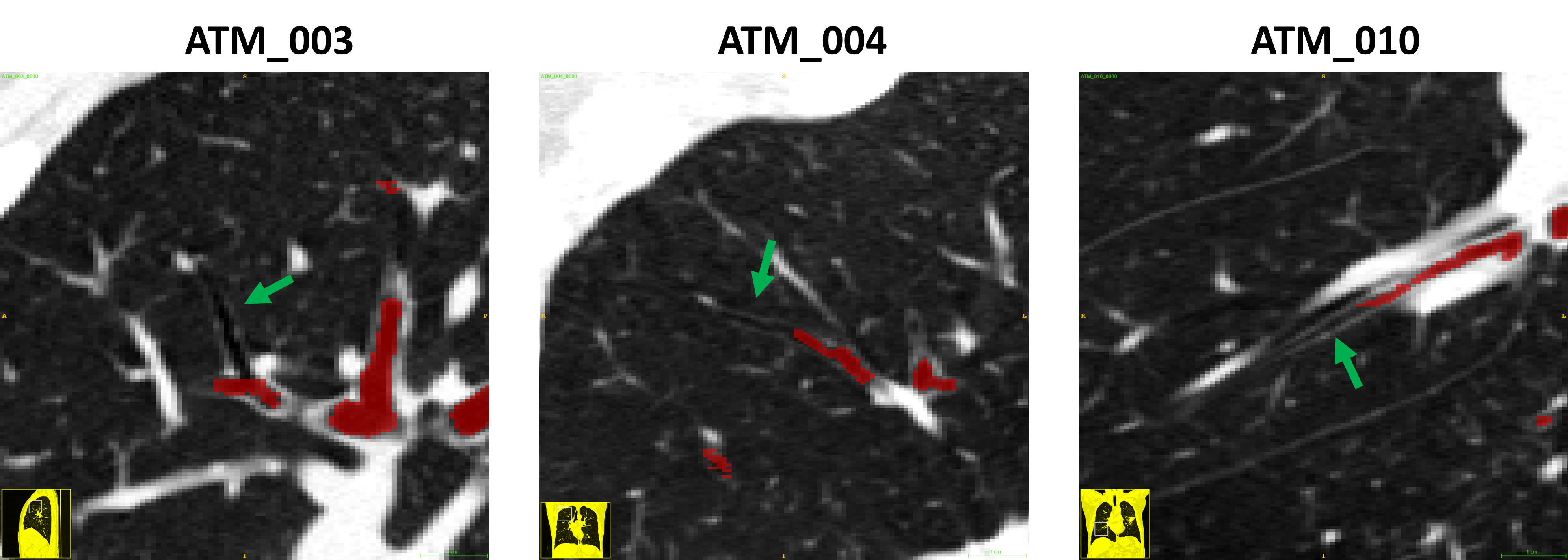}
\caption{Reference airway label of ATM'22 training set, where the airway tree is incomplete labeled (indicated by green arrows). 
}
\label{Fig:atm22_incomplete_demo} \vspace{-3mm}
\end{figure} 

\subsection{Comparative Evaluation on ATM’22 Dataset}
Quantitative results on ATM'22 testing dataset is reported in Table~\ref{table_atm}. Several conclusions can be drawn. First, our method (named DeepTree among ATM'22 participating teams) ranks 1st when using the weighted average score proposed by the organizers ( Equation 17 and Table 11 in~\cite{zhang2023multi}). Specifically, we achieves the best performance in BD, TLD and sensitivity with a markedly margin as compared to the second and third best performing methods in this relatively large-scale testing set. For instance, our BD, TLD and sensitivity improve those of \textbf{Timi} by 2.0\%, 2.4\% and 4.0\%, respectively; and are higher than those of \textbf{YangLab} by 3.4\%, 5.2\% and 3.5\%, respectively.  This indicates that a more complete airway tree can be segmented and extracted. Second, in model complexity, our method consumes the least amount of GPU memory among the top ten performing methods, offering more flexibility when deployed in clinical applications where the hardware capacity may be of concerns. Third, although our method exhibits a low precision score (87.9\%), we argue that this is mainly caused by the incomplete labeling issue in the testing set as discussed before. It is observed that reference labels provided by organizers are often incomplete in their training and validation set (Fig.~\ref{Fig:atm22_incomplete_demo} for examples). 

From the organizers' statement~\cite{zhang2023multi}, the labeling criterion is consistent for training, validation and testing. Hence, airway tree annotation in testing should yield similar labeling quality as the training and validation datasets. Our method is designed to segment the complete airway tree (although trained on the incomplete label), and the under-annotation issue in testing can notably impact the reported precision. Similar situations are found/discussed in evaluating BAS dataset. Our precision is 86.9\% when the original testing labels are used as ground truth but this performance score significantly improves to 98.4\% after careful curation of testing labels by three physicians.



\begin{figure*}[!t]
\centering
\includegraphics[width=0.95\textwidth]{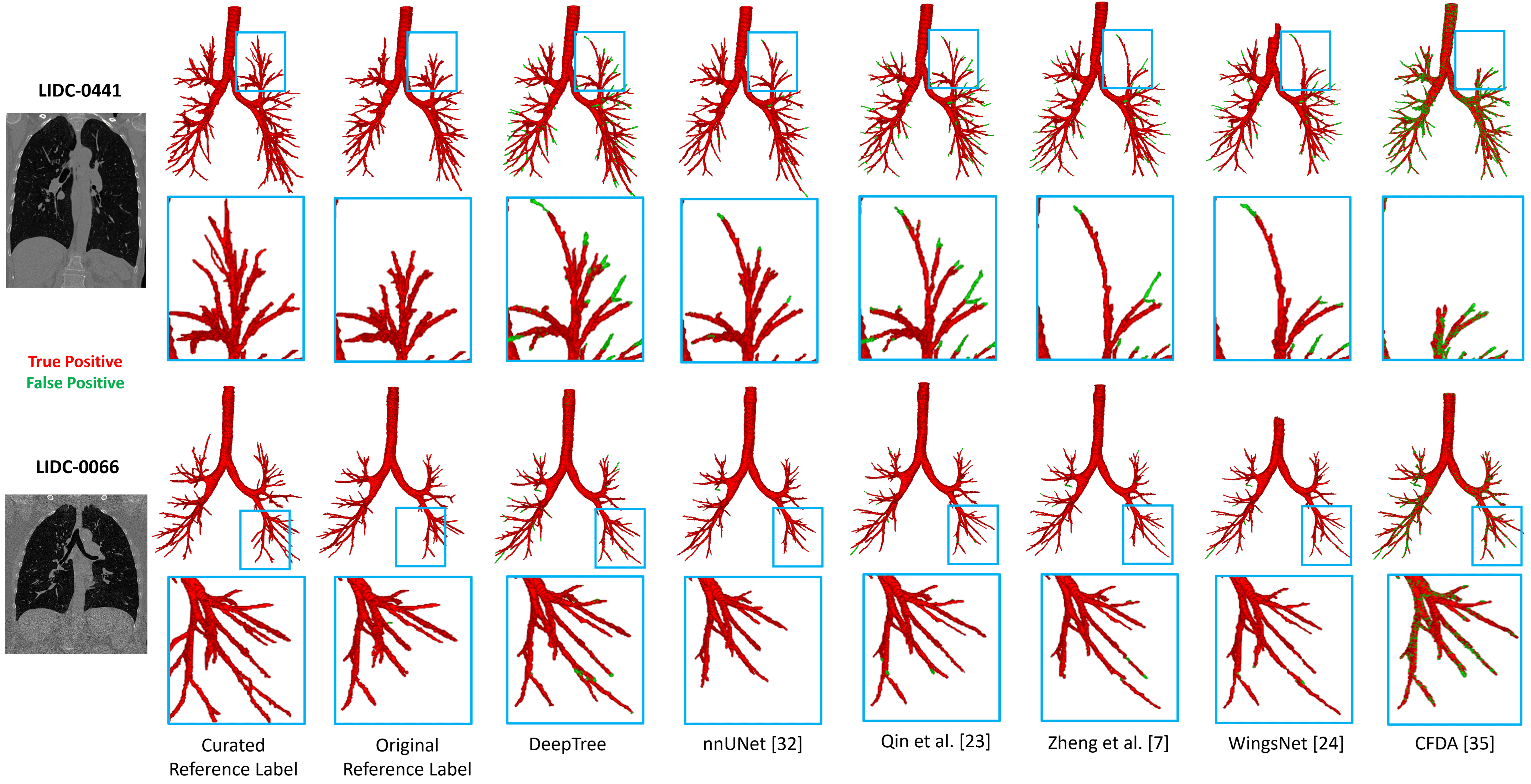}
\caption{Qualitative examples of airway tree segmentation results. 
}
\label{Fig:airwaySeg_quality}
\end{figure*}

\section{Conclusion}
In this work, we propose a new anatomy-aware multi-class segmentation method enhanced by  topology-guided iterative self learning. Motivated by the nature of airway anatomy, we formulate a simple yet highly effective anatomy-aware multi-class segmentation task to intuitively handle the severe intra-class imbalance of airway. To solve the incomplete labeling issue, we propose a tailored self-iterative learning scheme to segment towards the complete airway tree. For generating the pseudo-label of higher sensitivity (while retaining similar specificity) iteratively, we introduce a breakage attention map and design a topology-guided pseudo-label refinement method by connecting the breaking branches commonly existed in the initial pseudo-label. Extensively experiments are conducted on four datasets including two public challenges. The proposed method ranks 1st in the EXACT’09 challenge using the average score. It also ranks 1st in the ATM’22 challenge when using the weighted average score. Evaluation on two other datasets further validates that our proposed method can significantly improve over the leading approaches by extracting at least 7.5\% more detected tree length and 4.0\% more tree branches, while achieving similar precision levels.

\bibliographystyle{IEEEtran}
\bibliography{typeinst}

\end{document}